\documentclass[runningheads,a4paper]{llncs}

\usepackage{amsmath,amssymb}
\usepackage{bm}
\usepackage{cite}

\usepackage{graphicx}

\usepackage{url}
\urldef{\mailsa}\path|masuda@stat.t.u-tokyo.ac.jp|

\newcommand{\keywords}[1]{\par\addvspace\baselineskip
\noindent\keywordname\enspace\ignorespaces#1}

\newcommand{\EQ}{Eq.~}
\newcommand{\EQS}{Eqs.~}
\newcommand{\FIG}{Fig.~}

\begin{document}

\mainmatter  

\title{Application of Semidefinite Programming to Maximize the Spectral Gap\\ Produced by Node Removal}

\titlerunning{Semidefinite Programming for Maximizing the Spectral Gap}

%
%
\author{Naoki Masuda${}^{1,2}$%
\and Tetsuya Fujie${}^3$\and Kazuo Murota${}^1$}
\authorrunning{Naoki Masuda, Tetsuya Fujie, and Kazuo Murota}

\institute{${}^{1}$ Department of Mathematical Informatics,
The University of Tokyo,\\
7-3-1 Hongo, Bunkyo, Tokyo 113-8656, Japan\\
${}^2$
PRESTO, Japan Science and Technology Agency,\\
4-1-8 Honcho, Kawaguchi, Saitama 332-0012, Japan\\
${}^3$
Graduate School of Business,
University of Hyogo,\\
8-2-1 Gakuen-Nishimachi, Nishi-ku, Kobe 651-2197, Japan\\
\mailsa\\
\url{http://www.stat.t.u-tokyo.ac.jp/~masuda/}}

%
%

\toctitle{Semidefinite Programming for Maximizing the Spectral Gap}
\tocauthor{Authors' Instructions}
\maketitle

\begin{abstract}
The smallest positive eigenvalue of the Laplacian of a network is called the spectral gap and characterizes various dynamics on networks. We propose mathematical programming methods to maximize the spectral gap of a given network by removing a fixed number of nodes. We formulate relaxed versions of the original problem using semidefinite programming and apply them to example networks.
\end{abstract}

\section{Introduction}

An undirected and unweighted network (i.e., graph) on $N$ nodes is equivalent to an $N \times N$
symmetric
adjacency matrix $A=(A_{ij})$,
where $A_{ij}=1$ when nodes (also called vertices) $i$ and $j$ form 
a link (also called edge), and
$A_{ij}=0$ otherwise. We define the Laplacian matrix of the network by
\begin{equation}
L\equiv D-A,
\end{equation}
where
$D$ is the $N \times N$ diagonal matrix in which the $i$th diagonal
element is equal to $\sum_{j=1}^N A_{ij}$, i.e., the degree of node $i$.

When the network is connected, the eigenvalues of $L$
satisfy
\begin{equation}
\lambda_1=0 < \lambda_2 \le \cdots \le \lambda_N.
\end{equation}
The eigenvalue $\lambda_2$ is called spectral gap or algebraic connectivity
and characterizes various dynamics on networks including synchronizability
\cite{Almendral2007NJP,Arenas2008PhysRep,Donetti2006JSM},
speed of synchronization \cite{Almendral2007NJP},
consensus dynamics \cite{Olfati2007IEEE}, the speed of convergence of the
Markov chain to the stationary density 
\cite{Cvetkovic2010book,Donetti2006JSM}, and
the first-passage time of the random walk \cite{Donetti2006JSM}.
Because a large $\lambda_2$ is often considered to be desirable, e.g., for 
strong synchrony and high speed of convergence, 
maximization of $\lambda_2$ by changing networks under certain
constraints is important in applications.

In the present work, we consider the problem of maximizing
the spectral gap by removing a specified number, $N_{\rm del}$, of nodes from a given network.
We assume that an appropriate choice of $N_{\rm del}$ nodes keeps the
network connected.
A heuristic algorithm for this task in which nodes are sequentially removed is proposed in
\cite{Watanabe2010pre}. In this study, we explore a mathematical programming 
approach. We propose two algorithms
using semidefinite programming and numerically
compare their performance with that of the 
sequential algorithm proposed in \cite{Watanabe2010pre}. 

\section{Methods}

We start by introducing notations. 
First, the binary variable $x_i$ ($1\le i\le N$) 
takes a value of $0$ if node $i$ is one of the $N_{\rm
del}$ removed nodes and $1$ if node $i$ survives the removal.
Our goal is to determine $x_i$ ($1\le i\le N$) that maximizes $\lambda_2$
under the constraint
\begin{equation}
\sum_{i=1}^N x_i=N-N_{\rm del}.
\label{eq:constraint sum xi}
\end{equation}
Second, we define $\tilde{L}_{ij}$ as the $N \times N$ Laplacian matrix
generated by a single link $(i,j)\in E$, where $E$
is the set of links. In other words,
the ($i$,$i$) and ($j$,$j$) elements of $\tilde{L}_{ij}$ are
equal to 1, the ($i$,$j$) and ($j$,$i$) elements of $\tilde{L}_{ij}$
are equal to $-1$, and all the other elements of
$\tilde{L}_{ij}$ are equal to 0. It should be noted that
\begin{equation}
L=\sum_{1\le i<j\le N; (i,j)\in E}\tilde{L}_{ij}.
\label{eq:L sum}
\end{equation}
Third, $J$ denotes the $N \times N$ matrix in which all the $N^2$ elements are equal
to unity. Fourth, $E_i$ denotes the $N \times N$ diagonal matrix in which
the $(i,i)$ element is equal to unity and all the other $N^2-1$ elements are equal to 0.

After the removal of $N_{\rm del}$ nodes,
we do not decrease the size of the Laplacian. Instead, 
we remove $\tilde{L}_{ij}$
from the summation on the RHS of \EQ\eqref{eq:L sum}
if
node $i$ or $j$ has been removed from the network.
The Laplacian of the remaining network, if connected, has $N_{\rm del}+1$ zero
eigenvalues. The corresponding zero eigenvectors are given by
$\bm u^{(0)}\equiv 
(1 \cdots 1)^{\top}$ and $\bm e_i$, where $\top$ denotes the transposition,
$\bm e_i$ is the unit column vector
in which the $i$th element is equal to 1 and the other $N-1$ elements are
equal to 0, and $i$ is the index of one of the $N_{\rm del}$ removed nodes.

%
We formulate a nonlinear eigenvalue optimization
problem, which we call EIGEN, as follows:
\begin{equation*}
{\rm maximize} \; t 
%
%
\quad \mbox{ subject to \EQ\eqref{eq:constraint sum xi} and}
\end{equation*}
\begin{equation}
-t I + \sum_{i<j; (i,j)\in E}x_i x_j \tilde{L}_{ij} + \alpha J
+ \beta \sum_{i=1}^N (1-x_i) E_i \succeq 0,
\label{eq:constraint 1}
\end{equation}
and $x_i
\in \{0, 1\}\quad (1\le i\le N)$,
where $\succeq 0$ indicates that the LHS is a positive semidefinite matrix.
The positive semidefinite constraint \EQ\eqref{eq:constraint 1}
is derived from a standard prescription in semidefinite programming
for optimization of an extreme eigenvalue of a matrix.
Maximizing $t$ is equivalent
to maximizing the smallest eigenvalue of the matrix given by the
sum of the second, third, and fourth terms on the LHS
of \EQ\eqref{eq:constraint 1}.

Without the third and fourth terms on the LHS of
\EQ\eqref{eq:constraint 1}, the optimal solution would be trivially equal to
$t=0$ because the Laplacian of any network has 0 as the smallest
eigenvalue. Because $J=\bm u^{(0)}\bm u^{(0)\top}$, the third term
transports a zero eigenvalue to $\approx \alpha$.  We should take
a sufficiently large $\alpha>0$ such that the zero eigenvalue is shifted to a
value larger than the spectral gap of the remaining network, denoted
by $\tilde{\lambda}_2$.  This technique was introduced in
\cite{Cvetkovic1999LNCS} for solving the traveling salesman problem.

For each removed node $i$ (i.e., $x_i=0$),
the matrix
represented by the second term on the LHS of \EQ\eqref{eq:constraint 1}
has a zero eigenvalue associated with eigenvector $\bm e_i$.
The fourth term shifts this zero eigenvalue
to $\approx \beta$. Note that the fourth term disappears
for the remaining $N - N_{\rm
del}$ nodes because $x_i=1$ for the remaining nodes.
If the shifted eigenvalues are larger than $\tilde{\lambda}_2$,
the solution to the problem stated above
returns the $N_{\rm del}$ nodes whose removal maximizes $\tilde{\lambda}_2$.

The second term on the LHS of \EQ\eqref{eq:constraint 1} represents
a nonlinear constraint. To linearize
the problem in terms of the variables, 
we follow a conventional prescription 
%
%
to introduce auxiliary variables
\begin{equation}
X_{ij}\equiv x_i x_j,
\end{equation}
 where $1\le
i\le j\le N$
\cite{Grotschel1986jctsb,Lovasz1979IEEE,Lovasz1991SiamO}
(also reviewed in \cite{Goemans1997MP}).
If $x_i$ is discrete, $x_i (1-x_i)=0$ holds true.
Therefore, we require $X_{ii}=x_i^2 = x_i$. In the following discussion,
we use $x_i$ in place of $X_{ii}$.

We define the $(N+1) \times (N+1)$ matrix
\begin{equation}
Y\equiv \begin{bmatrix}
1 & \bm x^{\top}\\ \bm x & X
\end{bmatrix},
\label{eq:Y}
\end{equation}
where $\bm x \equiv (x_1\; \ldots \; x_N)^{\top}$,
the ($i,i$) element of the $N \times N$ matrix $X$ is equal to $x_i$, and
the ($i,j$) element ($i\neq j$) of $X$ is equal to $X_{ij}$.
By allowing $x_i$ and $X_{ij}$ ($1\le i< j\le N$) to take any 
continuous value between 0 and 1,
we define the relaxed problem named SDP1 as follows:
\begin{equation*}
{\rm maximize} \; t \quad \mbox{ subject to \EQ\eqref{eq:constraint sum xi} and}
\end{equation*}
\begin{align}
-t I + \sum_{i<j; (i,j)\in E}X_{ij} \tilde{L}_{ij} +& \alpha J
+ \beta \sum_{i=1}^N (1-x_i) E_i \succeq 0,
\label{eq:main constraint SDP1}\\
Y \succeq & 0.\label{eq:Y>=0}
%
\end{align}
Note that \EQ\eqref{eq:Y>=0} implies 
$0 \le x_i \le 1$ $(1\le i\le N)$ and that
SDP1 relaxes the original problem in that
$x_i$ and $X_{ij}$ are allowed
to take continuous values
while \EQ\eqref{eq:Y>=0} is imposed.
The method that we propose here for approximately maximizing
the spectral gap is to remove the $N_{\rm del}$ nodes
corresponding to the $N_{\rm del}$ smallest values among
$x_1$, $\ldots$, $x_N$ in the optimal solution of SDP1.
%

SDP1 involves $N(N+1)/2+1$ variables (i.e., $t$, $x_i$, and $X_{ij}$
with $i<j$). In fact, $X_{ij}$ for $(i,j)\notin E$ is free
unless \EQ\eqref{eq:Y>=0} is violated; it does not appear in the main
positive semidefinite constraint represented by \EQ\eqref{eq:main constraint
  SDP1}.  Because a given network is typically sparse, this implies
that there are many redundant variables in SDP1. To exploit the
sparsity and thus to save time and memory space, 
a technique based on matrix completion might be useful
\cite{Fukuda2000SiamO,Nakata2003MP}.  In this paper, however, we
propose another relaxation SDP2 for this purpose.

To linearize the second term on the LHS of
\EQ\eqref{eq:constraint 1}, we take advantage of four inequalities
$x_i x_j\ge 0$, $x_i(1-x_j)\ge 0$, $(1-x_i)x_j\ge 0$, and $(1-x_i)(1-x_j)\ge 0$ that must be satisfied for any link $(i, j)\in E$. By defining $X_{ij}\equiv x_i x_j$, as in the case of SDP1, we obtain the following four linear constraints \cite{Padberg1989MP}:
\begin{align}
X_{ij}\ge & 0,\label{eq:SDP2 linear constraint 1}\\
x_i-X_{ij}\ge & 0,\label{eq:SDP2 linear constraint 2}\\
x_j-X_{ij}\ge & 0,\label{eq:SDP2 linear constraint 3}\\
1-x_i-x_j+X_{ij}\ge & 0.\label{eq:SDP2 linear constraint 4}
\end{align}
SDP2 is defined by replacing \EQ\eqref{eq:Y>=0} by \EQS\eqref{eq:SDP2
  linear constraint 1}--\eqref{eq:SDP2 linear constraint 4}, where only
the pairs $(i,j)\in E$ are considered. Note that 
\EQS\eqref{eq:SDP2 linear constraint 1}--\eqref{eq:SDP2
  linear constraint 4} guarantee
$0\le x_i\le 1$ ($1\le i\le N$).
We remove the $N_{\rm del}$ nodes
corresponding to the $N_{\rm del}$ smallest values among
$x_1$, $\ldots$, $x_N$ in the optimal solution of SDP2.

Numerically, SDP2 is
much easier to solve than SDP1 for two reasons. First, the number of
variables is smaller in SDP2 than in SDP1. In SDP2, $X_{ij}$ is
defined only on the links, whereas in SDP1 it is defined for all the pairs $1\le
i<j\le N$. In sparse networks,
the number of variables is $O(N^2)$ for SDP1 and $O(N)$ for SDP2.
Second, the positive semidefinite constraint, which is much more 
time consuming to solve than a linear constraint of a
comparable size,
is smaller in SDP2 than in SDP1. While SDP1 and SDP2 share
the $N \times N$ positive semidefinite constraint
\eqref{eq:main constraint SDP1},
SDP1 involves an additional positive semidefinite constraint \eqref{eq:Y>=0} of size
$(N+1) \times (N+1)$.

To determine the values of $\alpha$ and $\beta$, we consider
the matrix
represented by the sum of the second, third, and fourth terms on the 
LHS of \EQ\eqref{eq:constraint 1}.
A straightforward calculation shows that the eigenvalues of 
this matrix are given by the
$N-N_{\rm del}-1$ positive eigenvalues of the Laplacian of the remaining 
network, ($N_{\rm
  del}-1$)-fold $\beta$, and $\beta+
\left[\alpha N - \beta \pm \sqrt{(\alpha N
    -\beta)^2 + 4 N_{\rm del} \alpha\beta} \right]/2$. For a
fixed $\beta$, we should select $\alpha$ to maximize
$\beta+
\left[\alpha N - \beta - \sqrt{(\alpha N
    -\beta)^2 + 4 N_{\rm del} \alpha\beta} \right]/2$, which is
always smaller than eigenvalue $\beta$.
We set
\begin{equation}
\alpha=\frac{\beta}{N}
\end{equation}
 to simplify the expression
of this eigenvalue to $\beta(1 - \sqrt{N_{\rm del}/N})$
while approximately maximizing this eigenvalue.

We have the following bounds for the optimal solution 
to the original problem.
We denote by $\tilde{\lambda}_2^{\rm opt}$ the optimal solution,
i.e., the maximum spectral gap with $N_{\rm del}$ nodes removed.
We denote by $\tilde{\lambda}_2^{\rm SDP}$
the smallest positive eigenvalue of the network 
obtained by the proposed method; the proposed method removes
the $N_{\rm del}$ nodes 
corresponding to the $N_{\rm del}$ smallest values of
$x_1$, $\ldots$, $x_N$ in the optimal solution of SDP1 or SDP2. 
Obviously, $\tilde{\lambda}_2^{\rm SDP}$ is a lower bound for
$\tilde{\lambda}_2^{\rm opt}$.
On the other hand, the optimal value, $\max t$, of SDP1 or SDP2 
serves as an upper bound for $\tilde{\lambda}_2^{\rm opt}$,
as long as the $\beta$ satisfies
$\tilde{\lambda}_2^{\rm opt} \leq \beta(1 - \sqrt{N_{\rm del}/N} )$.
This follows from the facts 
that the optimal value of EIGEN with such a $\beta$ value
coincides with $\tilde{\lambda}_2^{\rm opt}$ 
and both SDP1 and SDP2 are a relaxation of EIGEN.
We can summarize our observation as follows:
$\tilde{\lambda}_2^{\rm SDP} \leq \tilde{\lambda}_2^{\rm opt}
\leq \max t$. 

\section{Numerical results}

In this section, we
apply SDP1 and SDP2 to some synthetic and real networks.
We implement SDP1 and SDP2 using the free software package
SeDuMi 1.3 that runs on MATLAB 7.7.0.471 (R2008b) \cite{SeDuMi}.

We compare the performance of SDP1 and SDP2 with that of the optimal
sequential method, which is a heuristic method proposed in
\cite{Watanabe2010pre}.  In the optimal sequential method, we
numerically calculate the spectral gap for the network
obtained by the removal of one node; we
do this for all possible choices of a node to be removed.
Subsequently, we remove the node whose removal yields the
largest spectral gap. Then, for the remaining network composed of $N-1$
nodes, we determine the second node to be
removed in the same way. We repeat this procedure until $N_{\rm del}$ nodes have been removed.

\begin{figure}
\centering
\includegraphics[width=6cm]{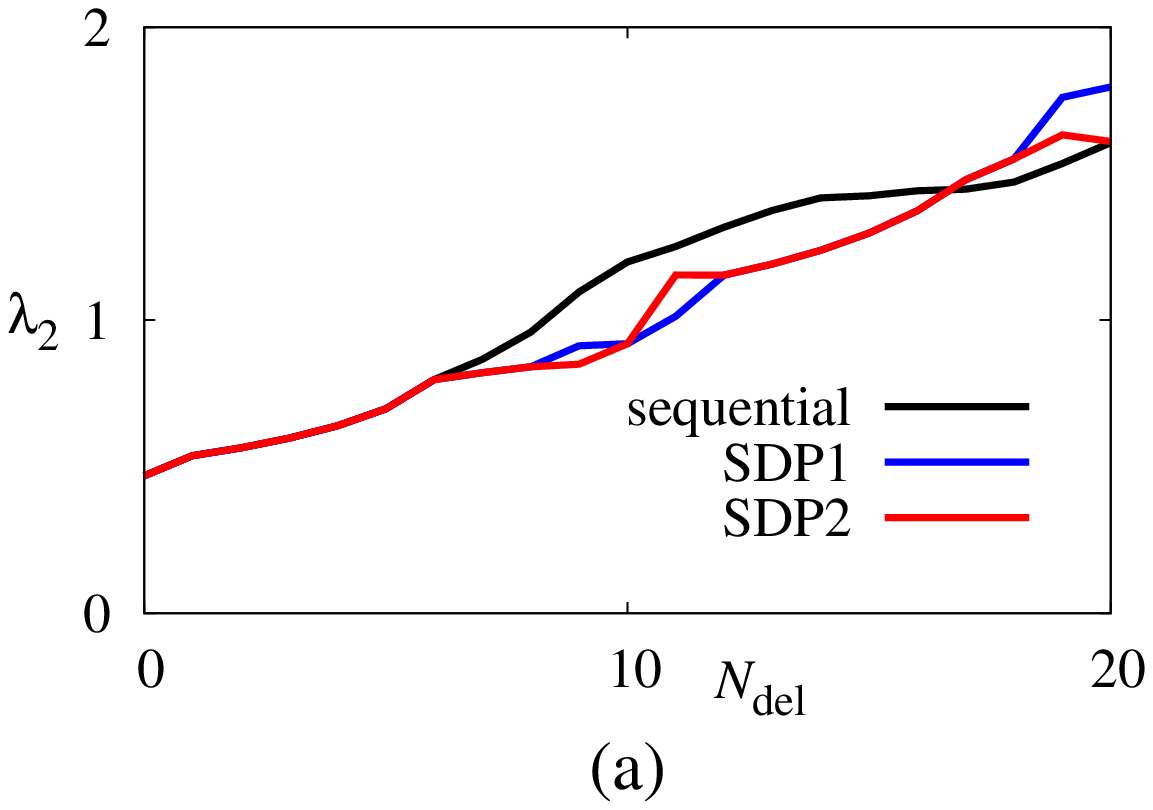}
\includegraphics[width=6cm]{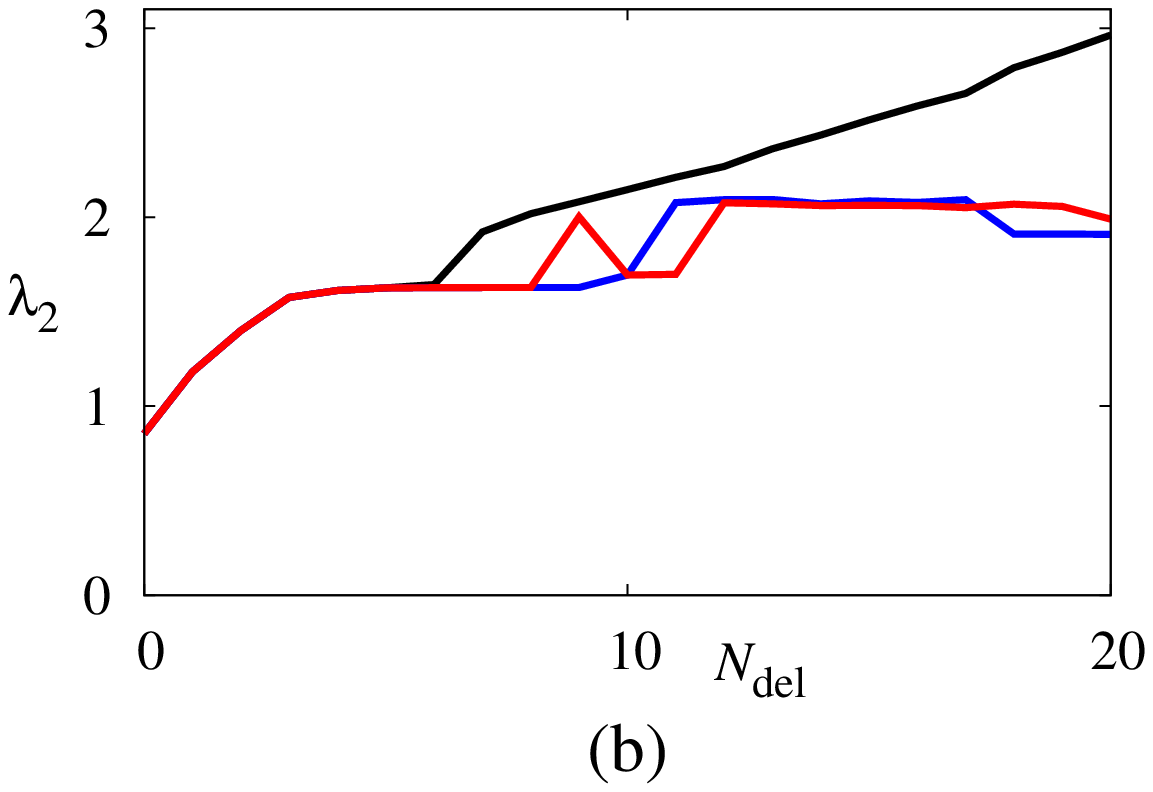}
\includegraphics[width=6cm]{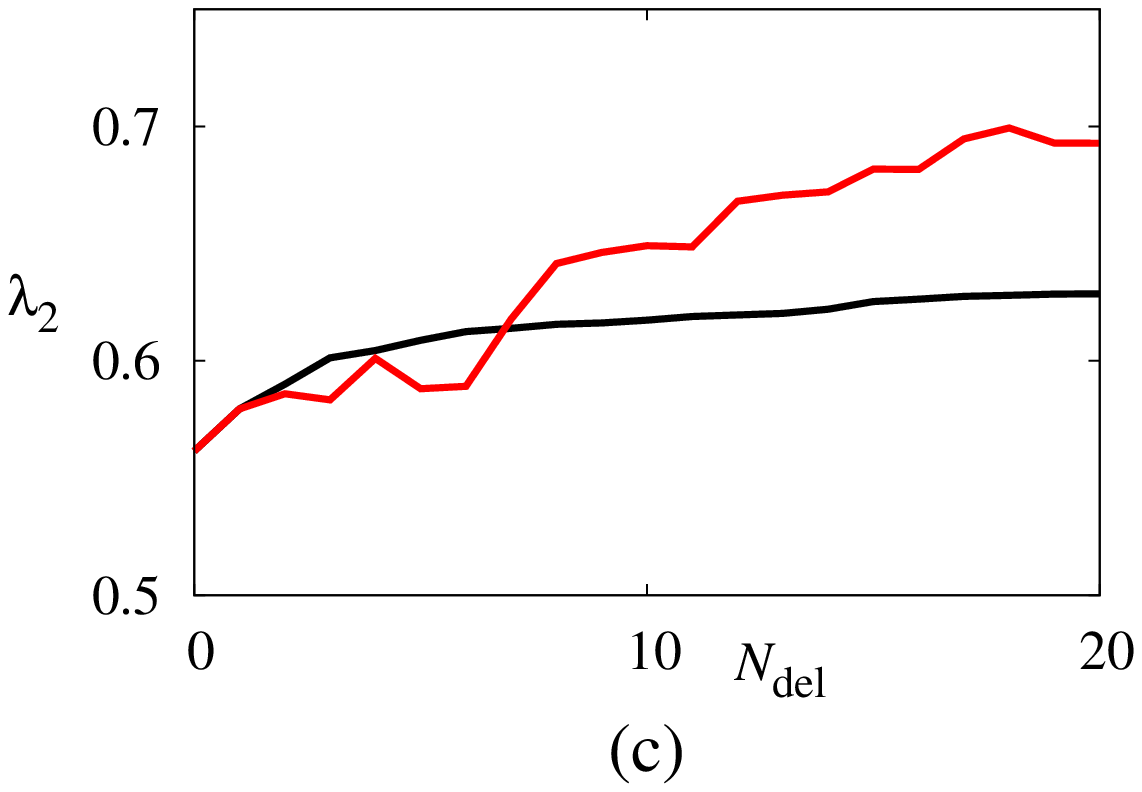}
\includegraphics[width=6cm]{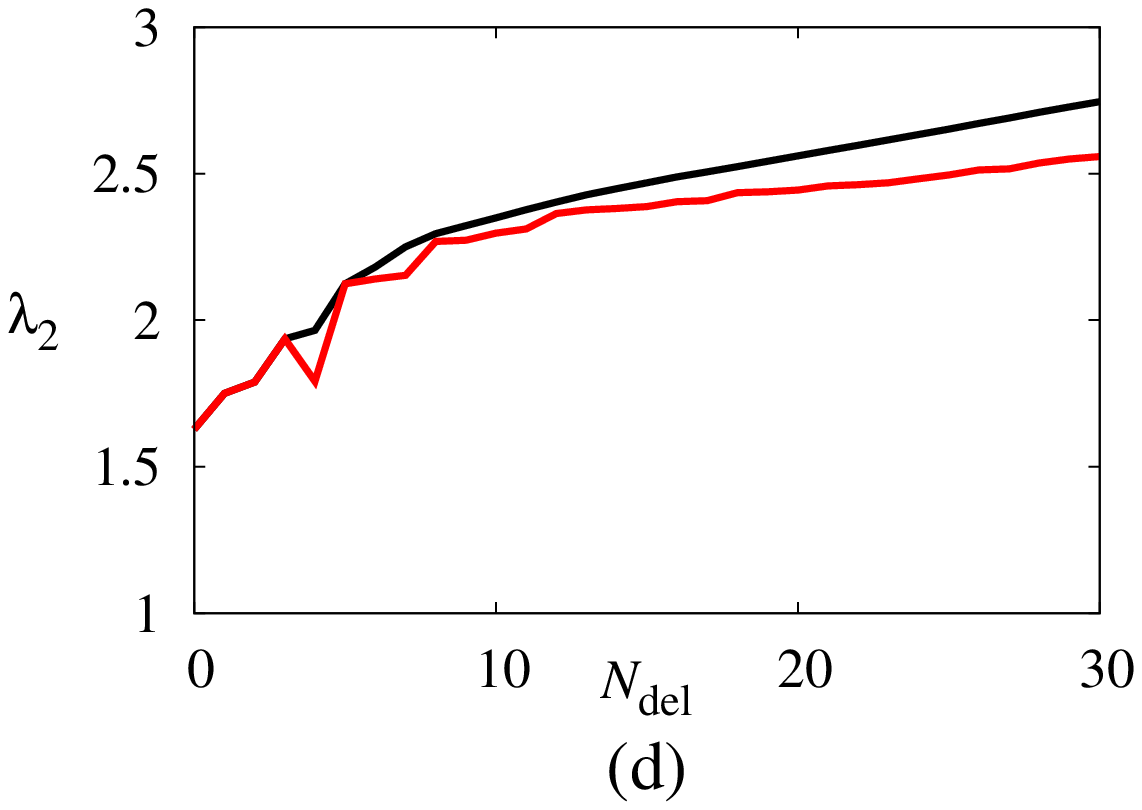}
\caption{Spectral gap as a function of the number of removed
nodes for four networks. (a) Karate club social network with $N=34$ nodes.
(b) Macaque cortical network with $N=71$ nodes.
(c) Barab\'{a}si--Albert scale-free network with $N=150$ nodes.
(d) \textit{C.~elegans} neural network with $N=279$
nodes.}
\label{fig:results}
\end{figure}

The first example network
is the well-known karate club social network, in which a node represents a member of the club and a link represents casual interaction between two members \cite{Zachary1977JAR}. The network has $N=34$ nodes and 78 links. We set $\beta=2$.
The spectral gaps obtained by the different node removal methods
are shown in \FIG\ref{fig:results}(a) as a function of $N_{\rm del}$.
Up to $N_{\rm del}=5$, the optimal sequential method yields the
exact solution, as do SDP1 and SDP2. For $N_{\rm del}\ge 6$, we could
not obtain the exact solution by the exhaustive search 
because of the combinatorial
explosion. For $7\le N_{\rm del}\le 16$, SDP1 and SDP2 perform worse than the
optimal sequential method. However, for $N_{\rm del}\ge 17$, both SDP1 and SDP2 outperform
the optimal sequential method. SDP1 and SDP2 found efficient combinations of removed nodes that the optimal sequential method could not find.

Second, we test the three methods against the largest connected component of the undirected and unweighted version of a macaque cortical network \cite{SpornsZwi04ni}. The network has
$N=71$ nodes and 438 links. We set $\beta=2$.
The spectral gaps obtained by the different methods
are shown in \FIG\ref{fig:results}(b).
Up to $N_{\rm del}=4$, the optimal sequential method yields the
exact solution, as do SDP1 and SDP2. For $N_{\rm del}\ge 5$, we could
not obtain the exact solution because of the combinatorial
explosion. For $N_{\rm del}\ge 5$, SDP1 and SDP2 perform worse than the
optimal sequential method. Consistent with the poor performance of SDP1 and SDP2,
the final values of $x_i$ ($1\le i\le N$)
are not bimodally distributed around 0 and 1 as SDP1 and SDP2 implicitly suppose.
The distribution is rather
unimodal except for the first three values of $x_i$ that are close to
0.  The ten values of $x_i$ when $N_{\rm del}=5$, in ascending
order, are as follows: $x_{33}=0.1086$, $x_{62}=0.1531$, $x_{53}=0.1589$,
$x_{1}=0.4813$, $x_{2}=0.5246$, $x_{8}=0.5591$, $x_{7}=0.6449$,
$x_{24}=0.7866$, $x_{51}=0.8749$, and $x_{63}=0.8931$ in SDP1, and
$x_{53}=0.000$, $x_{33}=0.145$, $x_{62}=0.177$, $x_{2}=0.585$,
$x_{1}=0.588$, $x_{8}=0.610$, $x_{7}=0.668$, $x_{24}=0.708$,
$x_{5}=0.738$, and $x_{4}=0.937$ in SDP2.

The third network is a network with $N=150$ nodes generated by the Barab\'{a}si--Albert scale-free network model \cite{Barabasi99sci}. The growth of the network starts with a connected pair of nodes, and each incoming node is assumed to have two links. The generated network has 297 links. We set $\beta=2$. For this and the next networks, SDP1 cannot be applied because $N$ is too
large. Therefore, we only compare the performance of SDP2 against the optimal sequential method. The results shown in 
\FIG\ref{fig:results}(c) indicate that SDP2 outperforms the optimal sequential method when $N_{\rm del}\ge 7$.

The fourth network is the largest connected component of the
\textit{C.~elegans} neural network \cite{wormatlas,Chen06pnas}. Two
nodes are regarded as being connected when they are connected by a
chemical synapse or gap junction. We ignore the direction and weight
of links. The network has $N=279$ nodes and 2287 links. We set
$\beta=2.5$. 
The results for SDP2 and the optimal sequential method are shown
in \FIG\ref{fig:results}(d).
Although the spectral gap gradually increases with
$N_{\rm del}$ for SDP2, SDP2
performs poorly as compared to the optimal sequential method for this example.

\section{Discussion}

We proposed a method to maximize the spectral gap using semidefinite
programming. The two proposed algorithms have a firmer
mathematical foundation as compared to the heuristic numerical method
(i.e., optimal sequential method). The proposed algorithms
performed better than the heuristic method for two networks especially for large
$N_{\rm del}$ and worse for the other two networks. For the former two networks, we could find the solutions in the situations in which the exhasutive search is computationally formidable. Up to our numerical efforts, our algorithms seem to be efficient for sparse networks.

We should be careful about the choice of $\beta$.
If $\beta$ is too large, SDP1 and SDP2 would
result in $x_i\approx N_{\rm del}/N$ ($1\le i\le N$). This is because
setting $x_i= N_{\rm del}/N$ ($1\le i\le N$)
makes the fourth term on the
LHS of \EQ\eqref{eq:constraint 1} equal
to $\beta \frac{N-N_{\rm del}}{N}I$, which increases all the
eigenvalues, including the spectral gap of the remaining network,
by $\beta \frac{N-N_{\rm del}}{N}$.
In contrast, if $\beta$ is smaller than $\tilde{\lambda}_2$,
SDP1 and SDP2 would maximize a false eigenvalue
originating from the fourth term
on the LHS of \EQ\eqref{eq:constraint 1}.

To enhance the performance of SDP1 and SDP2, it
may be useful to abandon the convexity of the problem. For example,
we could try replacing $(1 - x_i)$ in the fourth term by $(1 - x_i)^p$ and gradually
increase $p$ from unity. When $p>1$, the problem is no longer convex.
Accordingly, the existence of the unique solution and the convergence of
a proposed algorithm are not guaranteed.
Nevertheless, we may be able to track the optimal solution $\bm x$ by the
Newton method while we gradually increase $p$
(see p.5 and p.63 in \cite{BendsoeSigmundbook}).
%
%
%
%
An alternative extension is to add
$-p \sum_{i=1}^N x_i (1-x_i)$ to the objective function to be maximized (i.e.,
$t$).
When $p>0$, the convexity is violated. However, we may be able to adopt
a procedure similar to the method explained above, i.e.,
start with 
$p=0$ and gradually increase $p$ to track the solution by
the Newton method.



\subsubsection*{Acknowledgments.}

Naoki Masuda acknowledges the financial support of the Grants-in-Aid for Scientific
Research (no. 23681033) from MEXT, Japan.
This research is also partially supported by the Aihara Project, the FIRST
program from JSPS and by Global COE Program ``The research and training
center for new development in mathematics'' from MEXT.

\bibliography{../../directed/reducible/citations}









\end{document}